\documentstyle[aps,prb,epsf,twocolumn,epsfig,floats]{revtex}

\begin{document}
\draft



\wideabs{

\title{ Creation and pinning of vortex-antivortex pairs} %
\author{Sangbum Kim} %
\address{School of Computational Science, Florida State
University, Tallahassee, Florida 32306} %
\author{Chia-Ren Hu} %
\address{Center for Theoretical Physics, Department of Physics,
Texas A\&M University, College Station, Texas 77843} %
\date{March 14, 2006} %
\author{Malcolm J. Andrews} %
\address{Department of Mechanical Engineering, Texas A\&M
University, College Station, Texas 77843} %
\maketitle %
\begin{abstract} %

Computer modeling is reported about the creation and pinning of a
magnetic vortex-antivortex (V-AV) pair in a superconducting thin
film, due to the magnetic field of a vertical magnetic dipole above
the film, and two antidot pins inside the film. For film thickness
$= 0.1\xi$, $\kappa = 2$, and no pins, we find the film carries two
V-AV pairs at steady state in the imposed flux range $2.10\Phi_0 <
\Phi^+ < 3.0\Phi_0$, and no pairs below. With two antidot pins
suitably introduced into the film, a single V-AV pair can be stable
in the film for $\Phi^+ \ge 1.3\Phi_0$. At pin separation $\ge
17\xi$, we find the V-AV pair remains pinned after the dipole field
is removed, and, so can represent a 1 for a nonvolatile memory.
\end{abstract}

}  

\section{Introduction}

The electronic properties of superconducting devices are critically
influenced by the motion of vortices (Vs) within the superconductor
(SC). Pinning of Vs is responsible for critical current and magnetic
hysterisis in the behavior of a SC.~\cite{Tinkham} Any defects, such
as magnetic dots, twin boundaries~\cite{Crabtree-etal}, local spots
with lower critical temperature~\cite{Kato1994} $T_c$ or variable
thickness,~\cite{Coskun} can act as pinning centers.

Ferromagnetic particles, fabricated onto a superconductor, have many
effects of fundamental interest. Van Bael {\em et
al.}~\cite{VanBael-etal} considered a ferromagnetic dot array with
magnetic dipole moments parallel to the surface of superconducting
substrate. They showed that a flux lattice (FL) is pinned at the
opposite poles of the ferromagnetic dots (FDs), where flux quanta of
opposite signs are induced by the stray field. Nozaki {\em et
al.}~\cite{Nozaki-etal} investigated the effect of a spatially
modulated local field by a ferromagnetic dot array on a Nb film.
Since the magnetic field line from the ferromagnetic dots supresses
superconductivity where it penetrates, these magnetic dipoles (i.e.
FDs) can effectively create vortices and antivortices, and also have
a pinning effect on those vortices.

Milo\v{s}evi\'{c} and Peeters~\cite{MiloPeet} solved the
Ginzburg-Landau (GL) equations for the case of a ferromagnetic disk
(FDk) on top of a thin superconducting film. They found that, as
magnetic field lines penetrated a localized region of the film and
then returned to the outside, a giant antivortex (AV) formed under
the FDk, and several Vs located themselves in a ring outside the
region in the film below the FDk. Priour and Fertig~\cite{PriFert2}
performed a similar Ginzburg-Landau study for the vortex states in a
superconducting thin film, subject to the magnetic field of an FD
array, with the dipole moments oriented perpendicular to the film.
By varying the dipole moments, they changed the number density of
V-AV pairs, with the system going through various vortex
configurations that even break dot lattice symmetries.

Submicron-sized holes, or {\em antidots}, are well-known as strong
pins. Columnar defects created by the heavy-ion irradation technique
are known as efficient pinning centers, strongly enhancing the
critical current density.~\cite{Civale-etal} It has been
shown~\cite{MkrtShmidt,KhalShap} that a small antidot induces
negative potential energy around its center, attracting a V as a
singly-quantized flux line. Budzin~\cite{Budzin} has shown that for
large enough antidots, multiple flux quanta can be trapped.

Takezawa and Fukushima~\cite{TakeFuku} presented a simulation for
the pinning force on a V moving around a square antidot. They solved
the GL equations, and the dot size was varied to find the maximum
pinning force.  Priour and Fertig~\cite{PriFert} also performed a
numerical study of the behavior of Vs in the presence of an array of
antidots, in a thin superconducting film, using the GL equations.
They showed that when a V moves within a critical distance, $d_c$,
of an antidot, the V core deforms and extends to the antidot
boundary, and the associated supercurrent and magnetic flux spread
out and engulf the antidot.

Pearl~\cite{Pearl} solved the London equation for vortices moving in
a thin film, and showed that the vortices have a longer range for
their interaction force than in the bulk SC. This leads to the
concept of an {\em effective} penetration depth
$\lambda_{eff}=\lambda^2/2d$, which is much longer than the magnetic
penetration depth $\lambda$ in thin films where $d \ll \lambda$.

In this paper we report computer modeling of: (1) the creation of 2
and 3 V-AV pairs, but not one pair, in a thin {\em uniform}
superconducting film with a vertical magnetic dipole above it. The
dipole is made of two magnetic monopoles of strength $\pm q_m$,
separated by a distance $d_l$, with the positive monopole closer to
the film at a height $z_m$ above the film; (2) the creation and
pinning of a single V-AV pair in the film with two antidot pins
introduced in the film at appropriate locations and separation; and,
(3) the required V-AV separation to keep the pair pinned at two
antidots even after the dipole field is removed.~\cite{note1} The
motivation for this investigation is to investigate the scientific
feasibility of a non-volatile memory by using a magnetic dipole to
create a V-AV pair in a superconducting thin film, that relies on
two properly positioned pinning centers in the film that pin the
V-AV pair and prevent them from annihilating each other after
removal of the dipole. However, we do not attempt to analyze whether
such a memory device can compete with other existing memory devices
in cost and speed, we leave such concerns to future studies.

\section {The simplified Ginzburg-Landau model}

For a thin superconducting film with $d < \xi$ (the coherence
length) and $\lambda$, we neglect variations of all quantities in
the thickness direction of the film.~\cite{Chapman} After
nondimensionalization, the GL equations take the form:
\begin{equation}
\left(-i\nabla_{2D}-{\bf A}\right)^{2}\Psi = \Psi
\left(1-|\Psi|^{2}\right)\,,
\end{equation}
\begin{equation}
-\Delta_{3D}{\bf A} = (d /\kappa^{2})\delta(z){\bf j}_{2D}\,,
\end{equation} %
and
\begin{equation} %
{\bf j}_{2D} = {1\over 2 i}\left(
\Psi^{*}\nabla_{2D}\Psi-\Psi\nabla_{2D}\Psi^{*}
\right)-\left|\Psi\right|^{2}{\bf A}\,.
\end{equation} %
Distances are measured in units of $\xi$; the vector potential ${\bf
A}$ is in $\Phi_0/2\pi\xi$, where $\Phi_0\equiv h c/2e$ is the flux
quantum; the magnetic field is in $H_{c2} = \Phi_0/2\pi\xi^{2} =
\sqrt{2}\kappa H_{c}$, where $\kappa = \lambda/\xi$ is the
Ginzburg-Landau parameter. The London gauge $\nabla\cdot {\bf A} =
0$ has been employed for ${\bf A}$. The dimensionless
gauge-invariant free energy functional is then:
\begin{eqnarray}
G\left(\Psi,{\bf A}\right) & = & \int_{\Omega} \left(
-|\Psi|^{2}+{1\over 2}|\Psi|^{4}+\kappa^2 |\nabla\times {\bf A}-{\bf H}|^{2} \right. \nonumber \\
   &   & \left. \mbox{}+\left|\left({\nabla\over i}
         - {\bf A} \right)\Psi\right|^{2} \right) d\Omega\,,
\end{eqnarray}
where the $\Psi$-dependent terms are integrated inside the film
only. The first GL equation is solved by a relaxation
method~\cite{Kim-etal} for a pseudo-time-dependent equation
(${{\partial\Psi}\over {\partial t}}=-{{\delta G}\over
{\delta\Psi^{*}}}$), in a square sample with periodic boundary
conditions~\cite{DGR}. We are primarily interested in finding steady
states (note that the London gauge $\nabla\cdot {\bf A} = 0$ is a
good choice only at steady state). The relaxation method provides an
effective means to solve the nonlinear PDE, where the order
parameter changes in a pseudo-time, while the vector potential and
the supercurrent are updated simultaneously.

\begin{figure}[htb]
\begin{center}
\begin{picture}(120,120)
\put (5,20){\line(1,0){105}} \put (5,60){\line(1,0){105}} \put
(5,100){\line(1,0){105}}

\put (20,5){\line(0,1){105}} \put (60,5){\line(0,1){105}} \put
(100,5){\line(0,1){105}}

\put (10,40){\line(1,0){5}} \put (20,40){\line(1,0){5}} \put
(30,40){\line(1,0){5}} \put (40,40){\line(1,0){5}} \put
(50,40){\line(1,0){5}} \put (60,40){\line(1,0){5}} \put
(70,40){\line(1,0){5}} \put (80,40){\line(1,0){5}} \put
(90,40){\line(1,0){5}} \put (100,40){\line(1,0){5}} \put
(110,40){\line(1,0){5}}

\put (10,80){\line(1,0){5}} \put (20,80){\line(1,0){5}} \put
(30,80){\line(1,0){5}} \put (40,80){\line(1,0){5}} \put
(50,80){\line(1,0){5}} \put (60,80){\line(1,0){5}} \put
(70,80){\line(1,0){5}} \put (80,80){\line(1,0){5}} \put
(90,80){\line(1,0){5}} \put (100,80){\line(1,0){5}} \put
(110,80){\line(1,0){5}}

\put (40,10){\line(0,1){5}} \put (40,20){\line(0,1){5}} \put
(40,30){\line(0,1){5}} \put (40,40){\line(0,1){5}} \put
(40,50){\line(0,1){5}} \put (40,60){\line(0,1){5}} \put
(40,70){\line(0,1){5}} \put (40,80){\line(0,1){5}} \put
(40,90){\line(0,1){5}} \put (40,100){\line(0,1){5}} \put
(40,110){\line(0,1){5}}

\put (80,10){\line(0,1){5}} \put (80,20){\line(0,1){5}} \put
(80,30){\line(0,1){5}} \put (80,40){\line(0,1){5}} \put
(80,50){\line(0,1){5}} \put (80,60){\line(0,1){5}} \put
(80,70){\line(0,1){5}} \put (80,80){\line(0,1){5}} \put
(80,90){\line(0,1){5}} \put (80,100){\line(0,1){5}} \put
(80,110){\line(0,1){5}}

\put (120,40){\vector(0,1){20}} \put (120,40){\vector(0,-1){20}}
\put (122,40){$h_{y}$} \put (80,0){\vector(1,0){20}} \put
(80,0){\vector(-1,0){20}} \put (80,-8){$h_{x}$}

\put (64,64){P} \put (24,64){W} \put (104,64){E} \put (64,24){S}
\put (64,104){N}

\put (42,62){w} \put (82,62){e} \put (62,42){s} \put (62,82){n}

\put (60,60){\circle*{10}} \put (20,60){\circle*{10}} \put
(100,60){\circle*{10}} \put (60,20){\circle*{10}} \put
(60,100){\circle*{10}}

\put (40,60){\circle*{5}} \put (80,60){\circle*{5}} \put
(60,40){\circle*{5}} \put (60,80){\circle*{5}}
\end{picture}
\end{center}
\caption{The staggered grid arrangement for cell nodes P,E,W,N,S and
faces e,w,n,s.} \label{StaggeredGrid}
\end{figure}
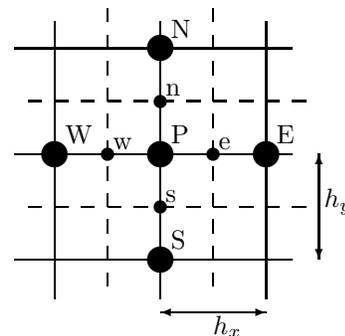

Using a link variable approach~\cite{Kaper,Kato1993} on a staggered
grid over $\Omega$ shown in Fig. \ref{StaggeredGrid}, the
semi-discrete equation is:

\scriptsize
\begin{eqnarray}
{\partial\Psi_P \over \partial t} & = & h_x h_y \left[ {{ d_w \left(
e^{i A_w h_x} \Psi_W - \Psi_P \right) + d_e \left( e^{-i
A_e h_x} \Psi_E - \Psi_P \right)} \over h_x^2} \right. \nonumber \\
      &    &  + \left. {{ d_s \left( e^{i B_s h_y} \Psi_S - \Psi_P \right) +
d_n \left( e^{-i B_n h_y} \Psi_N - \Psi_P \right) } \over h_y^2} \right] \nonumber \\
   &   & + \: h_x h_y d_p \left( 1 - |\Psi_P|^2 \right) \Psi_P\,,
\end{eqnarray} %
\normalsize

where $d=d(x,y)$ denotes the film thickness and $(h_x, h_y)$ the
spatial increments in $(x,y)$-directions. Note that the subscript of
each variable (except $x$ and $y$ in $h_x$ and $h_y$) denotes the
location at which it is defined on the grid (see Fig.
\ref{StaggeredGrid}).

The vector potential ${\bf A}$ is obtained by a fast Fourier
transform (FFT) technique.~\cite{Brigham} The 2-dimensional
supercurrent ${\bf j}_{2D} = ( j_x, j_y )$ on the grid is given by:

\scriptsize
\begin{eqnarray} %
j_{xe} & = & {1\over h_x} \left[ \left( \Phi_{P} \Theta_{E} -
\Theta_{P} \Phi_{E}\right) \cos \left(A_{e} h_{x}\right) -
\left(\Phi_{P} \Phi_{E} + \Theta_{P} \Theta_{E} \right) \sin
\left(A_{e} h_{x}\right) \right],  \nonumber  \\
j_{yn} & = & {1\over h_y} \left[ \left( \Phi_{P} \Theta_{N} -
\Theta_{P} \Phi_{N} \right) \cos \left(B_{n} h_{y}\right) -
\left(\Phi_{P} \Phi_{N} + \Theta_{P} \Theta_{N} \right) \sin
\left(B_{n}
h_{y}\right) \right]  \nonumber  \\
       &   &
\end{eqnarray}
\normalsize
where $\Phi$ and $\Theta$ are the real and imaginary
parts of $\Psi$, {\em i.e.} $\Psi = \Phi + i \Theta$.

 In this study, we have assumed $\kappa = 2$,
$h_x = h_y = 0.25$, and $\Delta t =$ 0.05 and 0.1 for spatial and
temporal discretization. For more details see~\cite{Kim}.

\section{Magnetic penetration in a film of constant thickness}

\begin{figure}
\centering
\includegraphics[width=8cm]{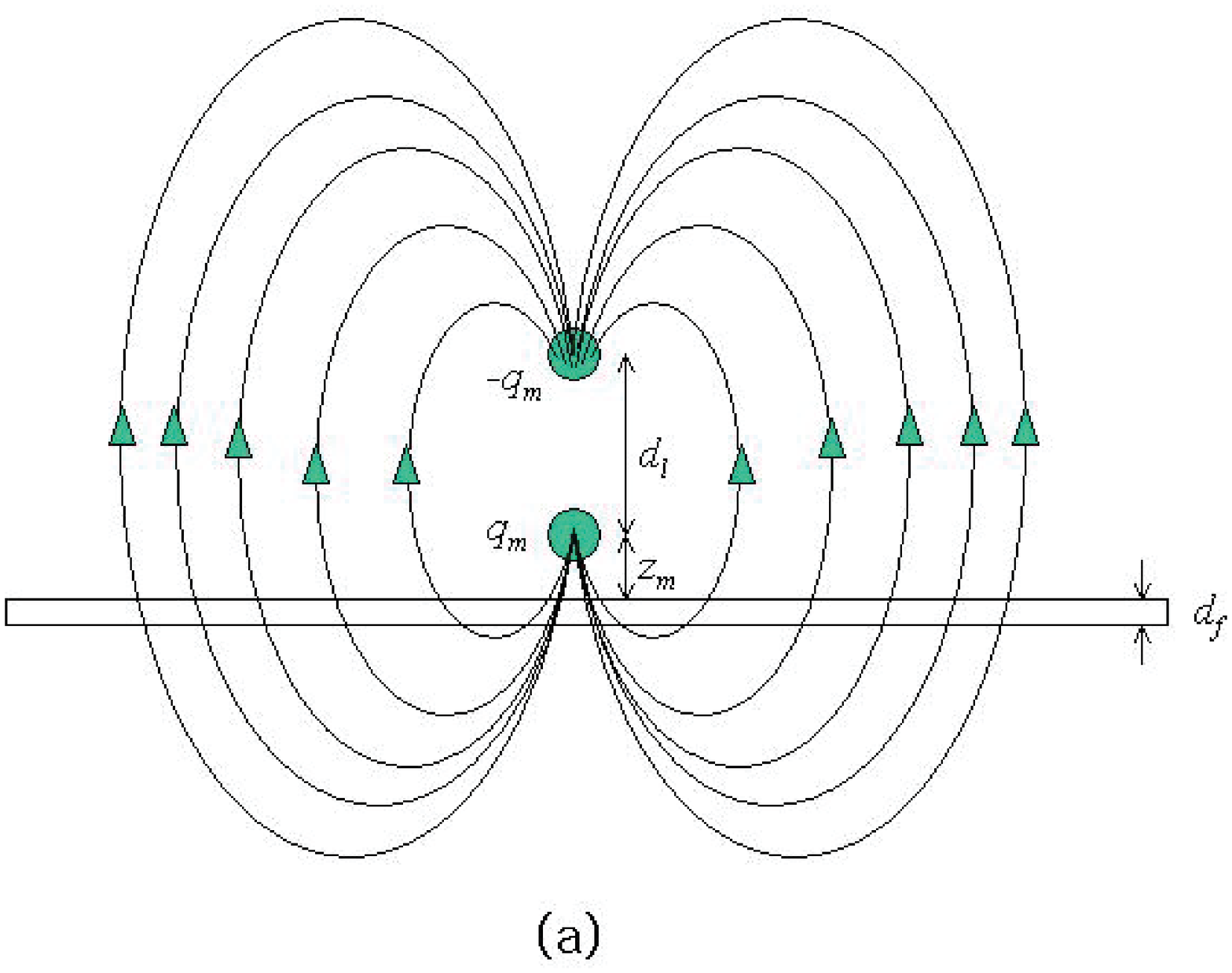}
\includegraphics[width=8cm]{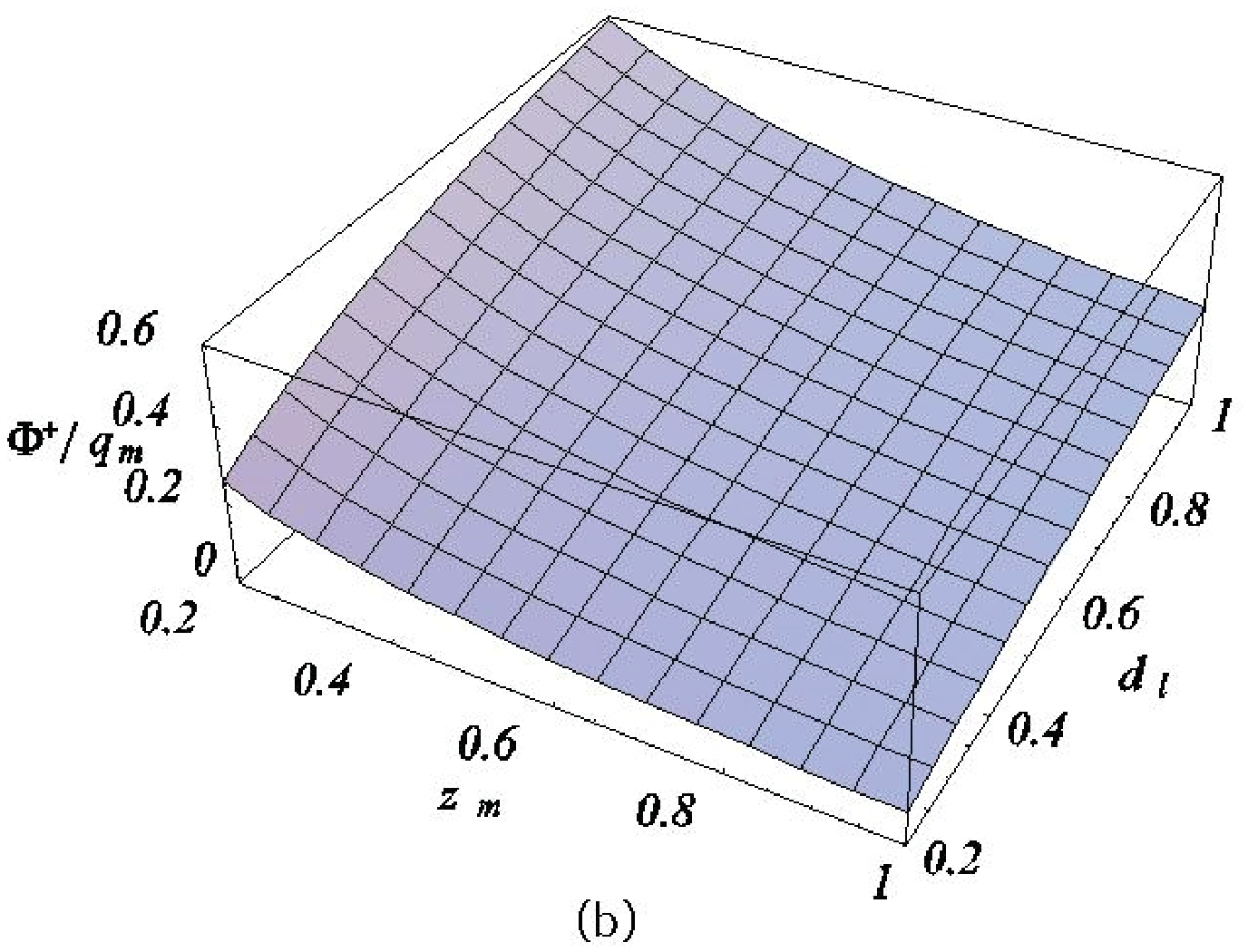}
\caption{(Color online) (a) A superconducting film under a magnetic
dipole. Stray field lines penetrating the film create
vortex-antivortex pair(s). They may be pinned at pinning centers
such as holes (antidots) made in the film in a proper configuration.
Since $d_{film} = 0.1 \xi$, the film is virtually transparent to
magnetic penetration in the thickness direction, and is ``soaked''
in the applied magnetic field. In the film, the induced magnetic
fields by the supercurrents incur only negligible change in the
applied magnetic field. (b) The 3-dimensional plot of $\Phi^+/q_m$
due to a {\em single} dipole. $\Phi^+/q_m$ decreases as $z_m$
increases, and increases as $d_l$ increases. } \label{system2}
\end{figure}

The film is subject to a non-uniform magnetic field from a magnetic
dipole above the film, and has pinning centers such as holes
(antidots) of various sizes and locations. Since $d_{film} = 0.1 \xi
\ll \lambda$, the film is virtually transparent to magnetic
penetration in the thickness direction, and is ``soaked'' in the
applied magnetic field (Fig.~\ref{system2}(a)).

Stray field from the magnetic dipole passing through the film
creates V-AV pairs. Since the Vs and AVs are topological excitations
of opposite quantum numbers, there must be an equal production of
them in the film. Field lines change polarity in the film so that
the total flux penetrating the film is zero.  We define the integral
of the downward part of the dipole field in the plane of the film,
i.e., $\Phi^+ = \int_{{\bf H}_M \cdot {\bf n} > 0} {\bf H}_M
\cdot{\bf n} d\sigma\,,$ where ${\bf n}$ is its downward normal, as
the imposed magnetic flux in the film by the magnetic dipole (see
Fig. \ref{system2}(a)). $\tilde\Phi^+\equiv\Phi^+/\Phi_0$ provides a
rough guide on how many V-AV pairs the dipole attempts to create in
the film.

For a single dipole, the magnetic flux can be written exactly as:
\begin{eqnarray}
\tilde\Phi^+ & = & {-1\over 2\pi} \int^{\rho_c}_0 2\pi\rho H_z
(\rho,
0) d\rho \nonumber  \\
             & = & -q_m \left( {\tilde z_m \over \sqrt{\rho_c^2
+ \tilde z_m^2}} - {\tilde z_m + d_l \over \sqrt{\rho_c^2 + (\tilde
z_m + d_l)^2}} \right),
\end{eqnarray}
where we have used the following expression for the z-component of
magnetic field:
\begin{eqnarray}
H_z (\rho, z) = & \nonumber  \\
 & \hspace{-.3in} -q_m \left( {z - \tilde z_m \over \left[\rho^2 + (z -
\tilde z_m)^2\right]^{3/2}} - {z - \tilde z_m - d_l \over
\left[\rho^2 + (z - \tilde z_m - d_l)^2\right]^{3/2}} \right).
\end{eqnarray}
 In the above $\rho^2 =
x^2 + y^2$ on the film, $\tilde z_m = z_m + {1\over 2} d_f$, and
$\rho_c$ is the radial location on the film where the z-component of
the magnetic field changes its polarity. It is given by the exact
expression $\rho_c^2 = \tilde z_m^{2/3} (\tilde z_m + d_l)^{4/3} +
\tilde z_m^{4/3} (\tilde z_m + d_l)^{2/3}$, when due to a single
dipole. But for a periodic array of dipoles, we have to evaluate it
numerically. (We make the thin film approximation by computing
$\tilde\Phi^+$ at the central layer of the film, which is at $z=0$.)
As $\rho_c$ increases the magnetic field lines reach farther
outside. However, to localize the field lines inside the
computational domain, we cannot use $z_m$ and $d_l$ that are too
large. Instead, we have to consider $z_m$ and $d_l$ that are much
smaller than the domain size. For this reason we have not
systematically explored their individual influence on the results
obtained. Instead, we have explored the effect of the most important
parameter $\tilde\Phi^+$ which can be achieved by varying $q_m$,
$z_m$, and/or $d_l$.

In addition, to consider a finite domain with periodic boundary
conditions, we have to use a periodic array the dipoles
($49\times49$ dipoles were superposed to simulate the periodic
array). The magnetic flux $\tilde\Phi^+$ calculated for the periodic
array of dipoles differs from that using a single dipole by less
than 5\%, in the range of $z_m$ and $d_l$ considered.

\begin{figure} %
\centering
\includegraphics[width=9cm]{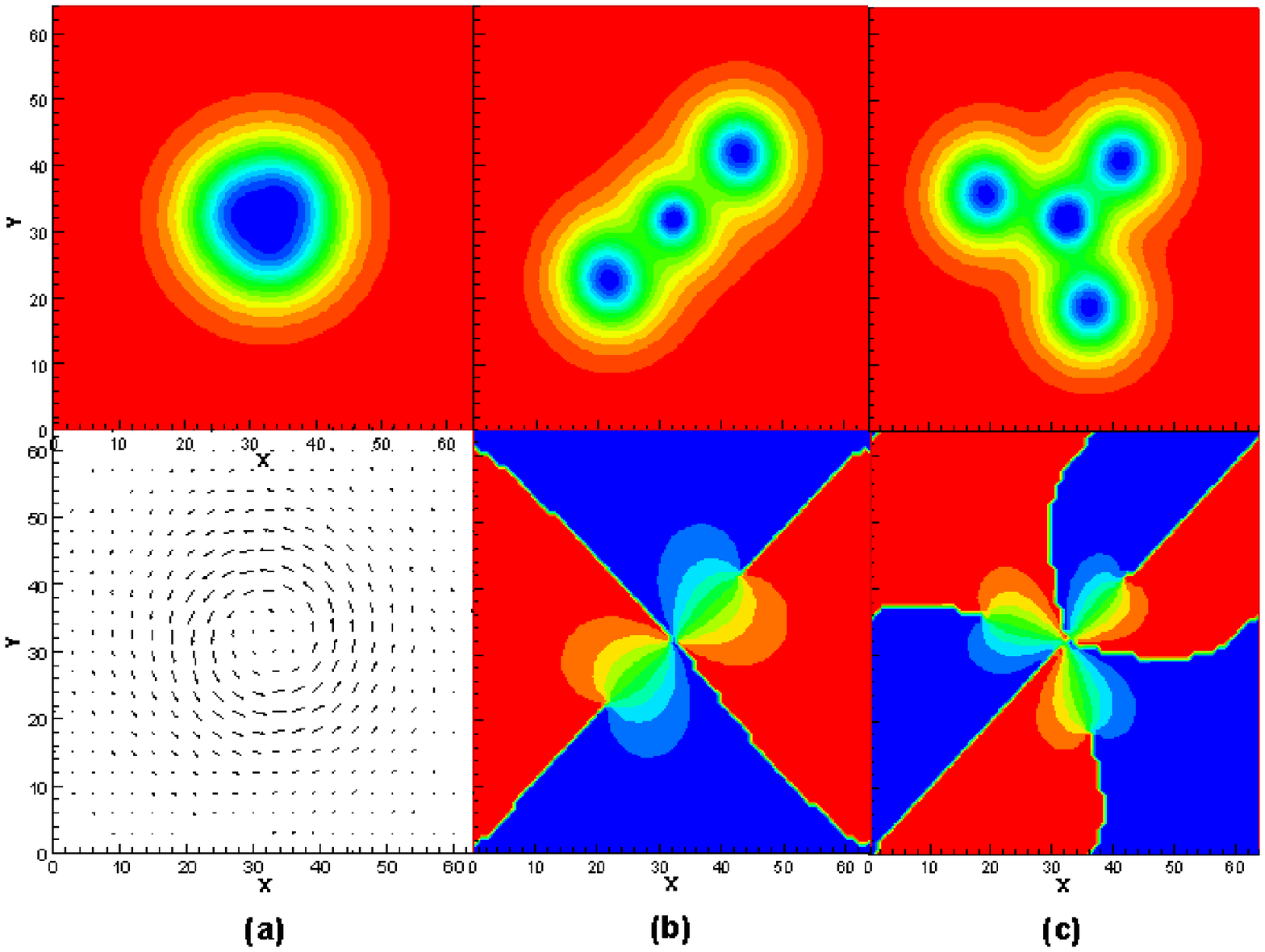}
\includegraphics[width=9cm]{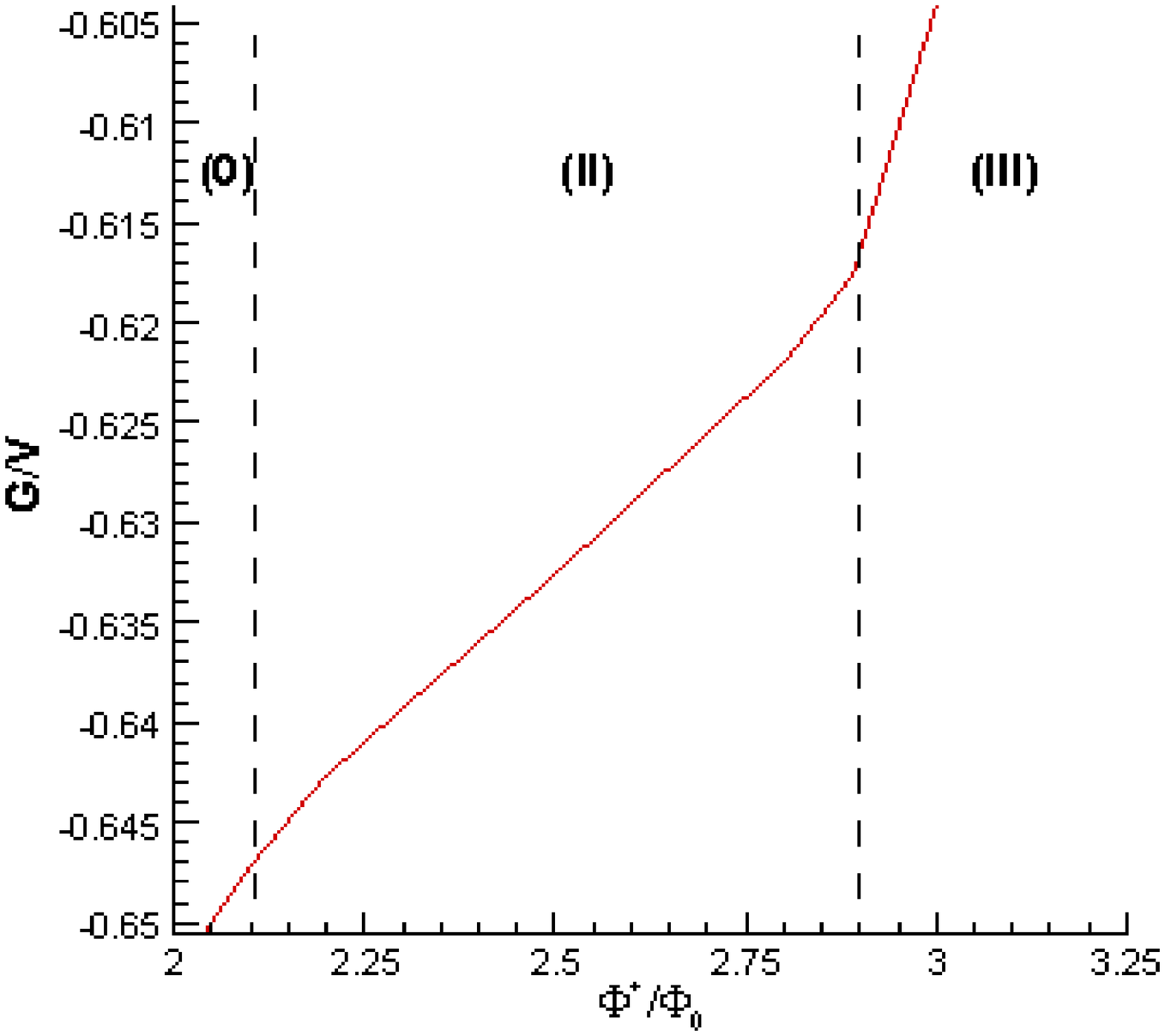}
\vspace{.1cm} \caption{\footnotesize (Color online) The
steady-states of magnetic penetration into a uniform film with no
pins under the external field from a magnetic dipole. The top is the
contour plots of the superelectron density (blue = 0, red = 1), and
the bottom is the corresponding phase plots (blue = 0, red = $2\pi$)
of the order parameter except the vector plot for supercurrent in
(a). Here $\tilde\Phi^+ = 2.05$ for (a), $\tilde\Phi^+ = 2.50$ for
(b), and $\tilde\Phi^+ = 3.0$ for (c). The film carries three V-AV
pairs in the steady state shown in (c), two pairs in (b), and no
pairs in (a). (d) shows the phase diagram of the steady state Gibbs
free energy ${G\over V} = {1\over V} \int_{A_{film}} \left( \kappa^2
{\bf A} \cdot {\bf j}_{2D} - {1\over 2} | \Psi
 |^4 \right) dA $. As $\tilde\Phi^+$ increases, the rising magnetic
penetration (vortices) also lifts G/V by increasing the
electromagnetic interaction energy and lowering the condensation
energy. The dashed lines separate regions of different number of
V-AV pairs: 0, 2, and 3 pairs (denoted by Roman numerals). Also, the
free energy curve steepens going from 2 to 3 V-AV pairs.}
\label{2pairband}
\end{figure}

Figure \ref{2pairband} shows magnetic penetration into a thin film
of thickness $d_f = 0.1$, with no antidot pins. The computational
domain is $16\xi \times 16 \xi$, with periodic extensions of this
geometry. We fix $z_m = 0.8\xi$, $q_m = 20 \Phi_0$, and vary $d_l$
to change $\Phi^+$. At $\tilde\Phi^+ = 3.0$ we find three V-AV pairs
[Fig. \ref{2pairband}(c)], and for $2.10 < \tilde\Phi^+ < 3.0$ the
film has two V-AV pairs [see Fig. \ref{2pairband}(b), for
$\tilde\Phi^+ = 2.50$].

The AVs accumulate near the center of the film, (i.e., directly below
the dipole), to form a giant AV, whereas the Vs spread outward symmetrically
because they repel each other and, although they are attracted to
the giant AV, they prefer to exist in the region where the magnetic
field points upward.

Plots of the phase of the order parameter give a clearer picture.
Thus, in Figs. \ref{2pairband} (a), (b) and (c) respectively, the
density plot is at the top and, for (b) and (c), the corresponding
phase plot is at the bottom, and the phase varies by $2\pi$ around
the core of each (singly-quantized) V. It varies by $4\pi$($6\pi$)
in the opposite direction around the core of a
doubly(triply)-quantized giant AV. The phase plots show that the Vs
and AVs are paired one to one.

However, no V-AV pairs are formed for $\tilde\Phi^+ \le 2.10$. In
Fig. \ref{2pairband} (a) ($\tilde\Phi^+ = 2.05$), the magnetic field
penetrates a large area in the film, just below the dipole. The area
near the center of the film is largely penetrated due to the high
value of the magnetic field around the center. At steady-state, no
vortex forms and the screening supercurrent (shown at the bottom of
Fig. \ref{2pairband}) is circulating only in one direction, as
required by the theory of London~\cite{Londons} for the Meissner
state under a dipole field. Fig. 3(d) is a phase diagram showing the
steady state Gibbs free energy ${G\over V} = {1\over V}
\int_{A_{film}} \left( \kappa^2 {\bf A} \cdot {\bf j}_{2D} - {1\over
2} | \Psi
 |^4 \right) dA $. As $\tilde\Phi^+$ increases, the rising magnetic
penetration (vortices) also lifts G/V by increasing the
electromagnetic interaction energy and lowering the condensation
energy. The dashed lines separate regions of different number of
V-AV pairs: 0, 2, and 3 pairs (denoted by Roman numerals). Also, the
free energy curve steepens going from 2 to 3 V-AV pairs.

Milo\v{s}evi\'{c} and Peeters~\cite{MiloPeet} suggested that the
number of V-AV pairs increase with $\tilde\Phi^+$ in a non-integral
increment of $\Delta\tilde\Phi^+ = 1.073$ when they used an FDk
dipole of radius $r_d = 1\xi$ and thickness $d_d = 1\xi$. In their
study, more V-AV pairs tend to be created as the radius of the FDk
decreases for a given dipole strength. Our dipole, consisting of a
magnetic monopole-antimonopole pair separated by a small distance,
corresponds to an extremely thin magnetic tip above the film and
produces a highly focused magnetic field. Thus, our
$\Delta\tilde\Phi^+ = 0.85$ between Fig. \ref{2pairband}(a) and (c)
agrees qualitatively with their conclusion, implying that it becomes
easier to create V-AV pairs as the lateral size of the dipole
decreases. Furthermore, in Ref. 7 the (dimensionless) dipole moment
$m/H_{c2} \xi^3 \approx 11 - 15$ is necessary to create 2 V-AV pairs
for $R_d = 3.5\xi$, whereas our dipole moment $q_m d_l \approx 5.2 -
8.3$ with $\tilde\Phi^+$ covering about the same percentage of the
film area, creates the same number of V-AV pairs. Thus, our point
dipole (magnetic tip) is about twice as effective at creating 2 V-AV
pairs than the FDk in Ref. 7. This we believe is because our dipole
field is highly non-uniformly distributed within the area of radius
$\rho_c$.

The magnetic dipole considered here generates a highly inhomogeneous
magnetic field in the superconducting film. The very strong downward
field, in a small region just below the dipole, holds all AV's to
form a giant AV. The much weaker upward field outside this region is
not effective at pulling the V's away from the giant AV. A mutual
repulsion between the V's also plays an important role in
determining the final V arrangement.~\cite{MiloPeet} Indeed, the
absence of such a repulsion in the case of a single V-AV pair may
account for the fact that we cannot obtain a single V-AV pair in any
$\tilde\Phi^+$ range. In Ref. 7, this lack of repulsive force was
compensated by the effect of the finite-size ferromagnetic disk.

Our computer modeling is based on a relaxation method with a
pseudo-time.~\cite{Kim-etal} Thus, computed transient states can
provide some qualitative information about the way the system
evolves in real time.
\begin{figure} %
\centering
\includegraphics[width=9cm]{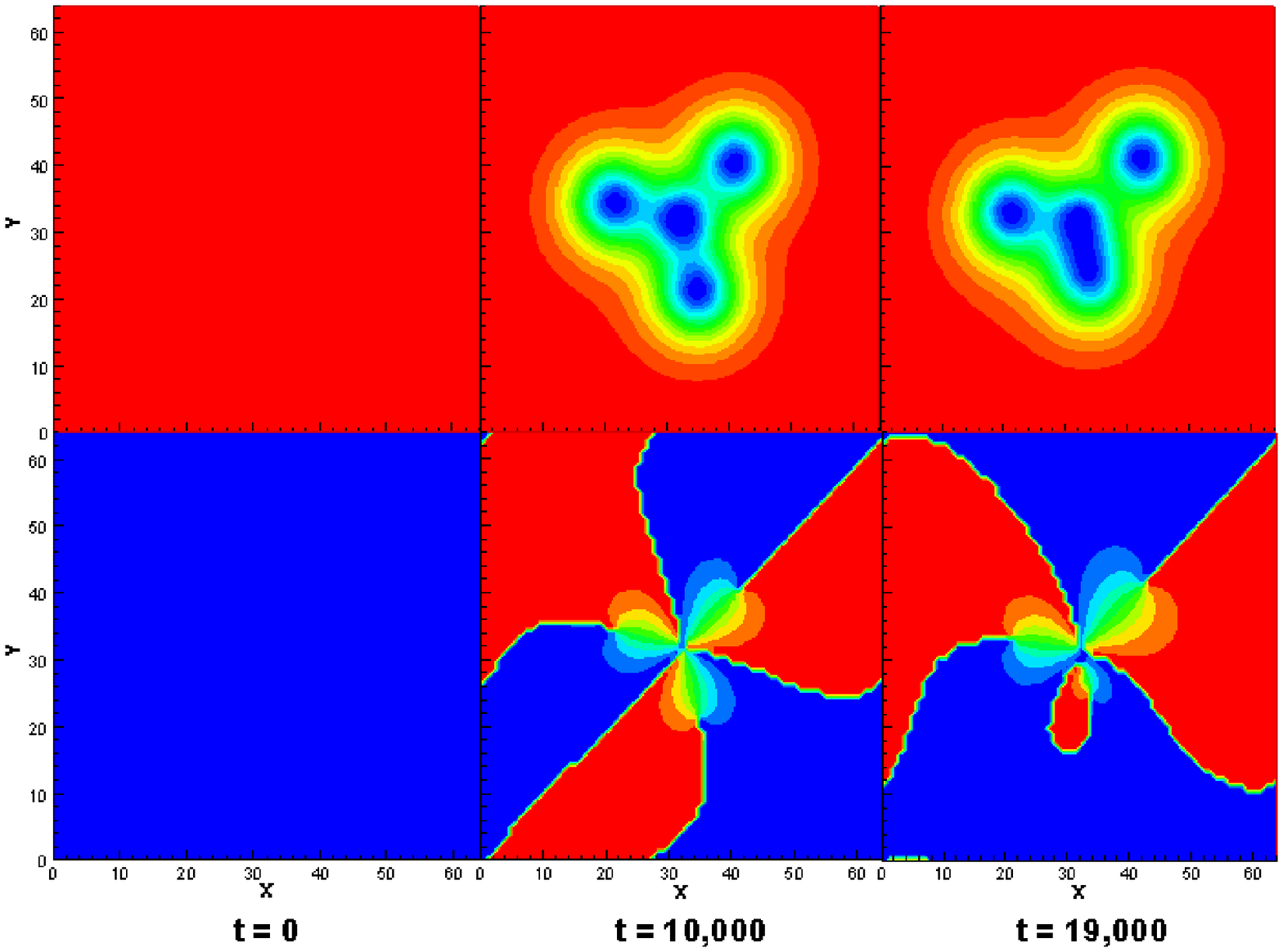}
\vspace{.2cm}
\includegraphics[width=9cm]{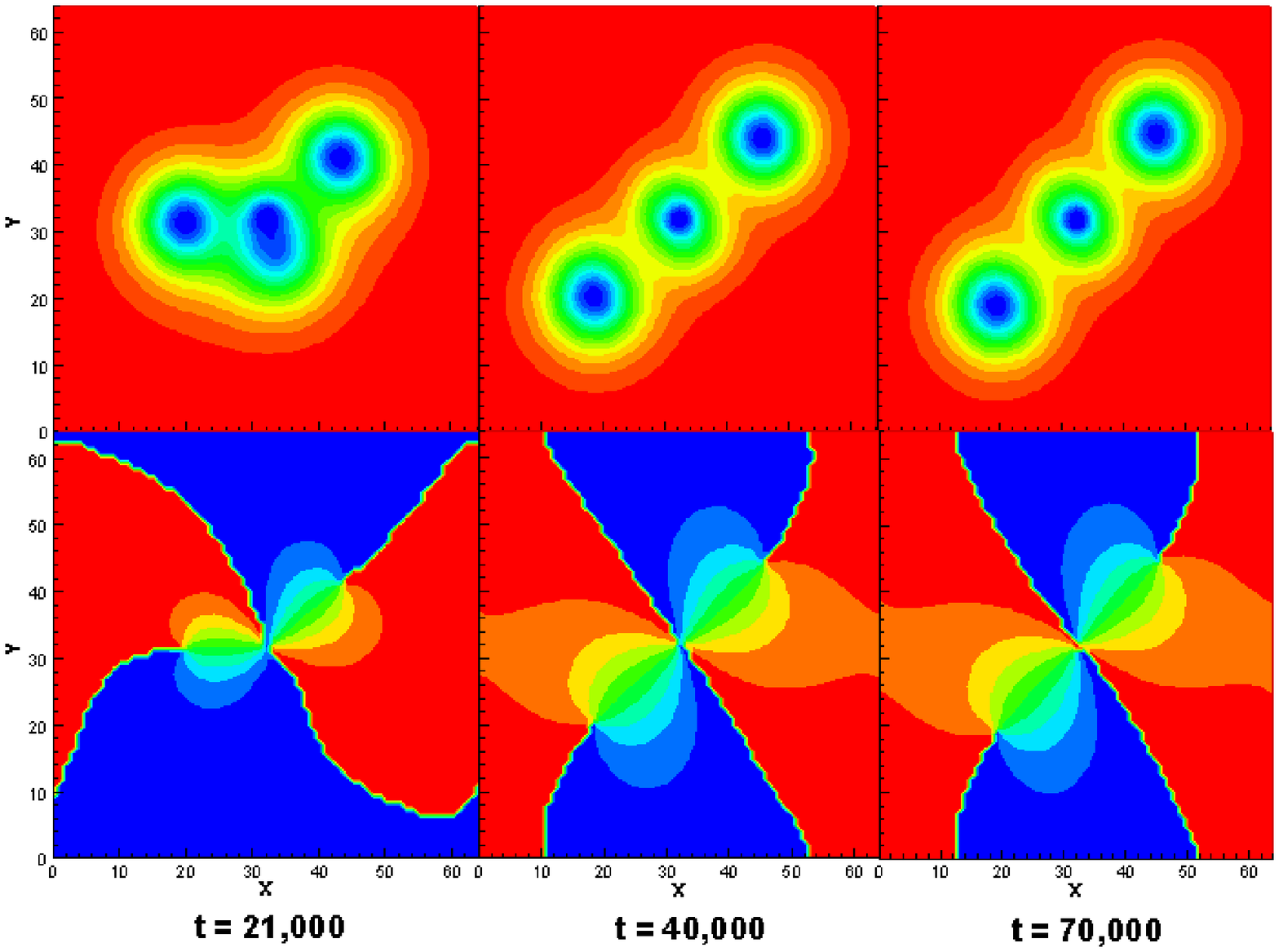}
\vspace{.2cm} \caption{(Color online) Pseudo-time sequence of
magnetic penetration into the film with no pins under the external
field from a magnetic dipole. At each time a pair of plots is shown,
with the density plot at the top and the corresponding phase plot at
the bottom. Here $\Phi^+ = 2.80$ and is in the interval $2.15 <
\Phi^+ < 3.0$ discussed in Fig.~\ref{2pairband}. Three V-AV pairs
initially appear in the film early (t=10,000), then one pair
annihilates itself (t=19,000 and 21,000). This process is more
discernible in the phase plot. Finally, the film carries two V-AV
pairs at steady state.} \label{partialann}
\end{figure}
In Fig.~\ref{partialann} two V-AV pairs are formed at $\Phi^+ =
2.80$. Initially, three V-AV pairs are created. Afterwards a pair of
V-AV annihilate each other and two pairs remain in the steady state.
This is more discernable in the phase plots (cf.
Fig.~\ref{partialann}.)

We find that for $2.20 < \tilde\Phi^+ < 3.0$, three V-AV pairs are
created first. Afterwards, one V-AV pair annihilates itself, leaving
only two pairs in the final steady state. The steady-state
configuration shown in Fig. \ref{2pairband}(b) belongs to such a
situation. (This annihilation has not been observed at $\tilde\Phi^+
= 2.20$, because the film carried two pairs throughout, suggesting
it was close to the lower boundary of the two V-AV pair region,
below which no V-AV pairs exist in the steady state.) Since our
relaxation method begins with full penetration of the dipole field,
a high peak field just below the dipole initially causes a larger
number (3) of V-AV pairs to be created. Eventually, as the Vs
separate from the AVs and move into a weaker upward field region,
the system settles for only two V-AV pairs over the given
$\tilde\Phi^+$ range. In addition, the vortices arrange themselves
diagonally at the farthest possible distance conforming to the given
sample geometry and the periodic boundary conditions.

The transient states discussed here are, of course, observable at
best in dynamic measurements only, since not only they are not
thermodynamically stable states, we believe there are not even any
metastable states involved at any intermediate stage. Some transient
states can be quite long lived though, when they are nearly
symmetric, such as the three V-AV pair state shown in Fig. 4 before
one V-AV pair eventually annihilate each other, with two remaining
V-AV pair then rearrange their positions to produce the final stable
state.

\section{Magnetic penetration in a film with pins}

When a pair of antidot pins are present in the film at suitable
positions, we find that one V-AV pair can be formed as low as
$\tilde\Phi^+ = 1.3$. Under this condition a pair of V-AV is created
and then separated, each of which is attracted to an antidot by its
pinning force. In the final state, an AV is pinned at an antidot
placed near the center of the film (P1), and a V is pinned at an
antidot some distance away along the diagonal (P2).

\begin{figure} %
\centering
\includegraphics[width=9cm]{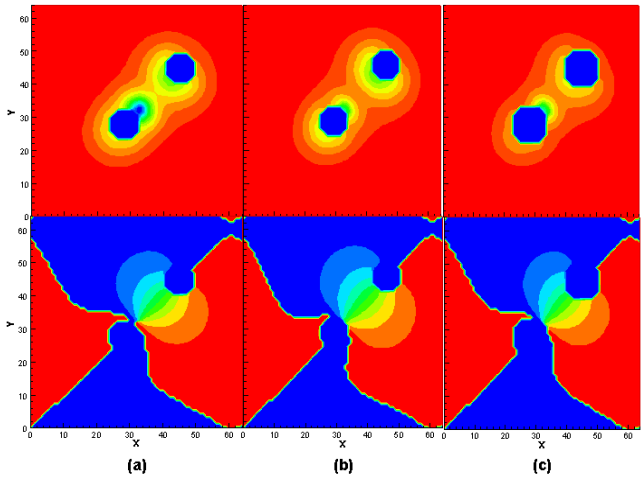}
\vspace{.3in} \caption{\footnotesize (Color online) For
$\tilde\Phi^+ = 1.90$ the steady-states of one V-AV pair creation in
a film with two pins, under the external field from a magnetic
dipole.  In (a) and (b), two pins, P1 and P2, are of the same size
and placed the same distance apart $(17/4)\sqrt{2}\xi$. The V-AV
pair are nicely pinned in (b). However, a small deviation from the
optimal pin position in (a) prevents full pinning (the AV stays near
the edge of P1 without being fully pulled in).  Changing the pins in
(a) to the bigger pins in (c) still does not achieve full pinning.}
\label{preferredloc}
\end{figure}

The effects of dipole field and pinning force on the final location
of the vortices is shown in Fig. \ref{preferredloc}. Here
$\tilde\Phi^+ = 1.90$, and for the two antidot pins we use square
holes clipped at their corners so they have octagonal shape. The
``diameter'' $d_p$ of these approximately round pins is defined to
be the edge length of the unclipped square hole. In each case of
Fig. \ref{preferredloc} (a) and (b), two pins with $d_p \cong 2\xi$
are placed at the same distance $(17/4)\sqrt{2}\xi$ apart along a
diagonal.

For (a), the center of the pin P1 is $\sqrt{2}\xi$ to the lower left
of the film center. At steady state, the AV stays around the edge of
P1 without being fully engulfed. However, for (b), where the center
of P1 is $(3/4)\sqrt{2}\xi$ to the lower left of the center, the AV
is completely absorbed into the pin. In Fig. \ref{preferredloc} (c),
the pins are bigger ($d_p \cong 2.5\xi$) and their positions are as
in (a). The AV is still not fully engulfed into P1, resulting in
incomplete (hence weaker) pinning. We conclude that optimal pinning
is sensitive to the precise location of the pins, and this
sensitivity is not removed by using a larger pin.

The dipole field strongly controls the positions of the V and the AV
of the pair. Since the film is soaked in the applied magnetic field
(Fig. \ref{system2}), the V-AV pair is simultaneously under the
influence of the dipole field and the pinning force. Thus, there
should exist an optimal configuration of pin locations to obtain the
best result for pinning. This optimum pin positioning is shown by
Fig. \ref{preferredloc} (b) (for the values of the parameters
chosen).

\section{Magnetic relaxation in a film with pins}

Due to the need for a larger computational domain, excessive
computational cost prohibits a complete simulation of magnetic
penetration followed by magnetic relaxation for larger V-AV
separation. Since penetration and relaxation are two distinct
processes from each other, we focus on relaxation to see if a V-AV
pair can be pinned at the given separation. To that end, we take a
single V-AV state from which the system can start to relax.

 Pinning of a
V-AV pair is more difficult when the dipole field is turned off. The
V-AV pair escapes from the pins and annihilates itself, even for the
pin distance of $(17/4)\sqrt{2}\xi$ of Fig. \ref{preferredloc} (b),
which is the optimal (and maximum) pinning distance we have obtained
in the presence of the dipole field of $\tilde\Phi^+ = 1.90$. Thus,
based on observations in previous figures, we have placed two pins
at diagonal positions (with varying distances), and, following Ref.
\ref{Kim-etal}, an {\em artificial} V-AV pair. (See
also~\cite{Schmid,Clem,Hu1972}.)

In several numerical experiments with random positions of the V-AV
pair as initial conditions, the steady states obtained are always
the same as if the Meissner-state initial condition is used. This
path-independence is expected for a thermodynamic equilibrium state.
Thus, we have been able to create initial conditions in which a V-AV
pair resides inside pins with arbitrary separation along the
diagonal. By increasing the distance between the pins (between the V
and the AV) we can reduce the attractive force between
them.~\cite{Pearl}

\begin{figure} %
\centering
\includegraphics[width=8cm]{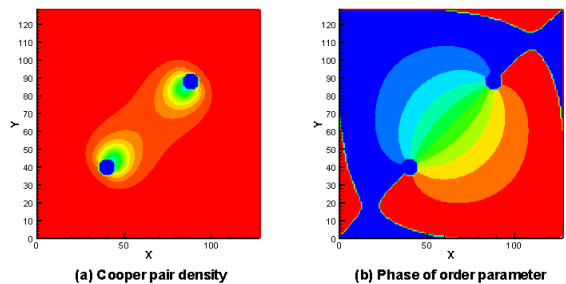}
\vspace{.2in} \caption{\footnotesize (Color online) The pins still
hold the V-AV pair after a long time, in a $32\xi \times 32\xi$
sample. We conclude that this computer modeling has reached
steady-state and that pinning of the V-AV pair has been achieved.}
\label{3dpinning}
\end{figure}

In Fig. \ref{3dpinning} the pin distance is about $12\sqrt{2}\xi$ in
a computational domain of size $32\xi \times 32\xi$. The system has
reached a steady-state without showing any further movement of the
V-AV pair from the pins. The pair initially attempted to escape the
pins with a tendency to annihilate itself, but it eventually
failed to emerge from the pins.

Thus, we have achieved our goals (2) and (3) of the introduction to
obtain a vortex state in which a V-AV pair is trapped in the pins,
and is unable to overcome the pinning force and annihilate itself.

Core deformation of the pair shown in Fig. \ref{3dpinning} is
characteristic of TDGL analysis (not observable in London analysis)
such as Ref. 14. So the pin distance $12\sqrt{2}\xi$ must be close
to the critical distance that pins the pair at the pinning spots.
(We observed annihilation of the pair at the pin distance of
$8\sqrt{2}\xi$, but did not systematically search for the precise
critical distance because each computation takes excessive
computational cost.)

If the V-AV pair remain pinned for a macroscopically long time after
the dipole field has been removed, then taken together they can
represent a 1 for nonvolatile data-storage, whereas the absence of a
V-AV pair at the pins can represent a 0. In this case the density of
vortices will be (1 V-AV pair)$/(32\xi)^2 \cong 4.7 GB/cm^2$,
assuming $\xi = 16 \:\: \AA$ for high temperature cuprate
superconductors.~\cite{note2} One concern is the possibility that
thermal fluctuations can overcome the energy barrier and de-pin the
V-AV pair, which would cause our memory idea to fail. Within the
Ginzburg-Landau theory we have estimated that the energy barrier of
a single V-AV pair is indeed slightly lower than the energy level of
thermal fluctuation ($k_B T$) for $Y Ba_2 Cu_3 O_{7-\delta}$ (YBCO)
at around 77 K assuming $d = 0.1\xi$. However, the reverse can be
true at somewhat lower temperatures (say, $0.5T_c$). Alternatively,
the barrier energy can be made larger by using a larger $d$,
allowing our idea to become valid at, say 77 K, but the present
theory is limited to $d \ll \xi$, to allow us to perform a two
dimensional calculation. Generalization to a three dimensional
calculation is conceptually simple, but expensive in computational
resources.

So far, only the write part of this new storage device has been
investigated, but we envision that reading can be achieved with a
scanning magnetic force probe, a scanning Hall probe, or a scanning
tunneling probe. Also we remark that by tilting the magnetic dipole
it should be easier to create a V-AV pair with the needed separation
for pinning after the dipole field is removed.

Our method to obtain precisely controlled vortex configurations may
provide for wider, and more versatile applications, e.g., Ref.
\ref{Gardner-etal} used a scanning SQUID microscope to create and
move a single V-AV pair. In Ref. \ref{Gardner-etal} the field coil
of the microscope was of $21 \mu m$ across, and hence applied a
force over hundreds of square microns. In contrast, our method
provides a controlled domain of submicron scale (only tens of
coherence lengths) even at this early research stage.

\section {Summary and Conclusion}

 A numerical study of the magnetization process in a thin type-II
superconducting film has been performed. The film is subject to a
non-uniform magnetic field created by a dipole above the film. The
dipole axis passes through the center of the film in the normal
direction to the film plane. For a film with constant thickness and
with no pin, it has been found that the film carries two V-AV pairs
at steady state in a range of the imposed magnetic flux
$\tilde\Phi^+$ ($2.10 < \tilde\Phi^+ < 3.0$), in units of flux
quantum. The 3-pair-to-2-pair transitions, in which a pair of V-AV
annihilates, have been observed in this range of $\tilde\Phi^+$. For
$\Phi^+ \leq 2.10$ no V-AV pair was created in the film. Two
antidots (holes) have been used to create a single V-AV pair in the
film, and a V-AV pair was created for lower magnetic fluxes to a
minimum $\Phi^+ = 1.3$. It was observed that the balance between the
dipole field and pinning force determines the optimal locations of
the V and AV. A simulation of the magnetization process in a sample
of size $32\xi\times 32\xi$ suggests that we are likely to achieve
pinning of the V-AV pair after removal of the dipole field with the
V-AV separation indicated by this sample size. Using this sample as
a template, the maximum density of pinned V-AV pairs achievable is
calculated to be about $4.7 GB/cm^2$.

\vspace{.1in} {\centerline {\bf ACKNOWLEDGEMENTS}} \vspace{.1in}

 Hu wishes to acknowledge partial support from the Texas Center
for Superconductivity and Advanced Materials at the University of
Houston. Andrews acknowledges support from the Texas A\&M University
through the Telecommunications and Informatics Task Force.


\begin{references}


\bibitem{Tinkham}  M. Tinkham, {\it Introduction to Superconductivity}
(McGraw-Hill Inc., New York, 1996).

\bibitem{Crabtree-etal} G. W. Crabtree, D. G. Gunter, H. G. Kaper,
A. E. Koshelev, G. K. Leaf, and V. M. Vinokur, Phys. Rev. B {\bf 61},
1446 (2000).

\bibitem{Kato1994}  R. Kato, Y. Enomoto, S. Maekawa, Physica C
{\bf 227} 387 (1994).

\bibitem{Coskun}  E. Coskun, Applied
Mathematics and Computation, {\bf 106} 31, (1999).

\bibitem{VanBael-etal} M. J. Van Bael, J. Bekaert, K. Temst, L.
Van Look, V. V. Moshchalkov, Y. Bruynseraede, G. D. Howells, A.
N. Grigorenko, S. J. Bending, and G. Borghs, Phys. Rev. Lett.
{\bf 86} 155, (2000).

\bibitem{Nozaki-etal} Y. Nozaki, Y. Otani, K. Runge, H. Miyajima,
B. Pannetier, J. P. Nozi\`{e}res and G. Fillion, J. Appl. Phys.
{\bf 79} (11) 8571, (1996).

\bibitem{MiloPeet} M. V. Milo\v{s}evi\'{c} and F. M. Peeters, J.
Low Temp. Phys. {\bf 130} 311, (2003).

\bibitem{PriFert2}  D. J. Priour, Jr. and H. A. Fertig, Phys. Rev.
Lett. {\bf 93} 057003 (2004).

\bibitem{Civale-etal} L. Civale, A. Marwick, T. Worthington, M. Kirk,
J. Thompson, L. Krusin-Elbaum, Y. Sun, J. Clem, and F. Holtzberg, Phys.
Rev. Lett. {\bf 67} 648, (1991).

\bibitem{MkrtShmidt} G. Mkrtchyan and V. Shmidt, Sov. Phys. JETP
{\bf 34} 195, (1972).

\bibitem{KhalShap} I. Khalfin and B. Shapiro, Physica C {\bf 207}
359, (1993).

\bibitem{Budzin} A. Budzin, Phys. Rev. B {\bf 47} 11416, (1993).

\bibitem{TakeFuku} N. Takezawa and K. Fukushima, Physica C
{\bf 290} 31 (1997).

\bibitem{PriFert}  D. J. Priour, Jr. and H. A. Fertig, Phys. Rev. B
{\bf 67} 054504 (2003).

\bibitem{Pearl} J. Pearl, Appl. Phys. Lett. {\bf 5} 65 (1964).

\bibitem{note1} In a thin superconducting film, the magnetic field induced
by supercurrents incurs negligible change in the applied magnetic
field. However, when the applied field is removed, this induced
field gives rise to an attractive V-AV interaction which is
comparable in magnitude to the current-induced attractive interaction
in the thin-film limit. Thus, pinning of a V-AV pair is harder to achieve
in a thin film than in a bulk SC.

\bibitem{Chapman}  S. J. Chapman, Z. angew. Math. Phys. {\bf 47} 410 (1996).

\bibitem{Kim-etal} S. Kim, C.-R. Hu and M. J. Andrews, Phys. Rev. B,
{\bf 69}, 094521 (2004).\label{Kim-etal}

\bibitem{DGR} M. \hspace{-.05in} Doria, J. \hspace{-.05in} Gubernatis, D.
\hspace{-.05in} Rainer, Phys. Rev. B {\bf 39} 9573 (1989).

\bibitem{Kaper}  H. G. Kaper and M. K. Kwong, J. Comp. Phys.
{\bf 119} 120 (1995).

\bibitem{Kato1993}  R. Kato, Y. Enomoto, S. Maekawa, Phys. Rev. B
{\bf 47} 8016 (1993).

\bibitem{Brigham} E. O. Brigham, {\it The Fast Fourier Transform and
its Applications} (Prentice-Hall, Inc., Upper Saddle River, NJ
1988).

\bibitem{Kim}  S. Kim, {\it A Numerical Study of Steady-State Vortex
Configurations and Vortex Pinning in Type-II Superconductors}, Ph.D.
dissertation, Department of Mechanical Engineering, Texas A\&M
University, College Station, TX (2004).

\bibitem{Londons} F. London and H. London, Physica {\bf 2} 341
(1935); F. London, Act. Sci. et Ind., No. 458 (1937), Paris.

\bibitem{Schmid} A. Schmid, Phys. kondens. Materie {\bf 5} 302 (1966).

\bibitem{Clem} J. R. Clem, J. Low Temp. Phys. {\bf 18} 427 (1975).

\bibitem{Hu1972} C.-R. Hu and R. S. Thompson, Phys. Rev. B {\bf 6}
110 (1972).

\bibitem{note2} The effective Ginzburg-Landau parameter
$\kappa_{eff} = {\lambda_{eff} \over \xi} = {\lambda^2 / d \over
\xi} = {\xi \over d} \kappa^2$, for $d=0.1\xi$ is $\kappa_{eff} =
40$. Since our sample size is only 32 $\xi$, any $\lambda$ or
$\lambda_{eff}$ which is much larger than this sample size is not
important. Since our $\lambda_{eff}$ is already somewhat larger than
the sample size, it gives large $\kappa$ behavior for the sample
size studied here. Since $\lambda_{eff}$ determines the
characteristic scale of V-AV in thin films as shown by Pearl, we
believe the results apply to the electromagnetic interaction of V
and AV in thin superconductor films with higher $\kappa$. For
example, it is reported that polycrystalline $Y Ba_2 Cu_3
O_{7-\delta}$ has $\kappa$ = 51 at 77 K and 47 at 0 K [A. Bourdillon
and N. X. Tan Bourdillon, {\em High Temperature Superconductors:}
{\em Processing and Science} (Academic Press, 1994)].

\bibitem{Gardner-etal} B. W. Gardner, J. C. Wynn, D. A. Bonn, R. Liang,
W. N. Hardy, J. R. Kirtley, V. G. Kogan, and K. A. Moler, Appl.
Phys. Lett. {\bf 80} 1010 (2002).\label{Gardner-etal}


\end{references}
\end{document}